\documentclass[12pt]{article}
\usepackage{psfig,amsmath}
\newcommand{\avida}{{\bf Avida}}

\begin{document}
\begin{flushright}
\small \noindent Proc. Nat. Acad. Sci. USA {\bf 97} (2000) 4463-4468
\end{flushright}
\vskip 0.5cm
\centerline{\LARGE\bf \sf Evolution of Biological Complexity}
\vskip 0.5cm
\centerline{\large \bf \sf  Christoph  Adami$^\dagger$, 
Charles Ofria$^\star$\protect\footnote{Present address: Center for Microbial
    Ecology, Michigan State University, Lansing, MI 48824}
and Travis C. Collier$^\S$}
\vskip 0.5cm
\centerline{\it $^\dagger$Kellogg Radiation Laboratory 106-38 and 
$^\star$Beckman Institute 139-74}
\centerline{\it California Institute of Technology, Pasadena, CA 91125} 
\vskip 0.2in
\centerline{\it  $^\S$Division of Organismic Biology,
Ecology, and Evolution,}
\centerline {\it University of California Los Angeles, Los Angeles, CA 90095}
\vskip 0.5cm

{\bf   In order to make a case for or against a trend in the evolution of
  complexity in biological evolution, complexity needs to be both
  rigorously defined and measurable.  A recent information-theoretic
  (but intuitively evident) definition identifies genomic complexity
  with the amount of information a sequence stores about its
  environment. We investigate the evolution of genomic complexity in
  populations of digital organisms and monitor in detail the
  evolutionary transitions that increase complexity. We show that
  because natural selection forces genomes to behave as a natural
  ``Maxwell Demon'', within a fixed environment genomic complexity is
  forced to increase.
}

Darwinian evolution is a simple yet powerful process that requires
only a population of reproducing organisms in which each offspring has
the potential for a heritable variation from its parent.  This
principle governs evolution in the natural world, and has gracefully
produced organisms of vast complexity. Still, whether or not
complexity increases through evolution has become a contentious
issue. Gould \cite{gould96} for example argues that any
recognizable trend can be explained by the ``drunkard's walk'' model,
where ``progress'' is due simply to a fixed boundary condition.
McShea~\cite{mcshea96} investigates trends in the evolution
of certain types of structural and functional complexity, and finds
some evidence of a trend but nothing conclusive. In fact, he concludes
that ``Something may be increasing. But is it complexity?''  Bennett
\cite{bennett95}, on the other hand, resolves the issue by {\it fiat},
defining complexity
as ``that which increases when self-organizing systems organize
themselves''. Of course, in order to
address this issue, complexity needs to be both defined {\em and}
measurable. 

In this paper, we skirt the issue of structural and functional
complexity by examining {\em genomic complexity}. It is tempting to
believe that genomic complexity is mirrored in functional complexity
and vice versa. Such an hypothesis however hinges upon both the
aforementioned ambiguous definition of complexity and the obvious
difficulty of matching genes with function.  Several developments
allow us to bring a new perspective to this old problem.  On the one
hand, genomic complexity can be defined in a consistent
information-theoretic manner (the ``physical''
complexity~\cite{adami99}), which appears to encompass intuitive
notions of complexity used in the analysis of genomic structure and
organization~\cite{britten71}.  On the other hand, it has been shown
that evolution can be observed in an artificial
medium~\cite{adami98,lenski99}, providing a unique glimpse at
universal aspects of the evolutionary process in a computational
world. In this system, the symbolic sequences subject to evolution are
computer programs that have the ability to self-replicate via the
execution of their own code. In this respect, they are computational
analogs of catalytically active RNA sequences that serve as the
templates of their own reproduction. In populations of such sequences
that adapt to their world (inside of a computer's memory), noisy
self-replication coupled with finite resources and an information-rich
environment leads to a growth in sequence length as the digital
organisms incorporate more and more information about their
environment into their genome~\cite{fn0}. These populations allow us
to observe the growth of physical complexity explicitly, and also to
distinguish distinct evolutionary pressures acting on the genome and
analyze them in a mathematical framework.

If an organism's complexity is a reflection of the physical complexity
of its genome (as we assume here) the latter is of prime importance in
evolutionary theory. Physical complexity, roughly speaking, reflects
the number of base pairs in a sequence that are functional. As is well
known, equating genomic complexity with genome length in base
pairs gives rise to a conundrum (known as the C-value paradox) because
large variations in genomic complexity (in particular in eukaryotes)
seem to bear little relation to the differences in organismic
complexity~\cite{cavalier85}.  The C-value
paradox is partly resolved by recognizing that not all of DNA is
functional; that there is a {\em neutral} fraction that can
vary from species to species. If we were able to monitor the
non-neutral fraction, it is likely that a significant increase in this
fraction could be observed throughout at least the early course of
evolution. For the later period, in particular the later Phanerozoic
Era, it is unlikely that the growth in complexity of genomes is due
solely to innovations in which genes with novel functions arise {\it
  de novo}. Indeed, most of the enzyme activity classes in mammals,
for example, are already present in prokaryotes~\cite{dixon64}.
 Rather, gene duplication events leading to
repetitive DNA and subsequent diversification~\cite{britten69} 
as well as the evolution of gene regulation
patterns appears to be a more likely scenario
for this stage. Still, we believe that the Maxwell Demon mechanism
described below is at
work during all phases of evolution and provides the driving force
toward ever increasing complexity in the natural world. 

\vskip 0.25cm 
\noindent {\bf Information Theory and Complexity} 
\vskip 0.25cm
Using information theory to understand evolution and the information
content of the sequences it gives rise to is not a new undertaking.
Unfortunately, many of the earlier attempts (e.g.,
Refs.~\cite{schroedinger45,gatlin72,wiley82})
confuse the picture more than
clarifying it, often clouded by misguided notions of the concept of
information~\cite{fn1}.  An (at times amusing) attempt to make sense of
these misunderstandings is Ref.~\cite{collier86}.

Perhaps a key aspect of information theory is that information cannot exist
in a vacuum, that is, information is {\em physical}~\cite{landauer91}.
This statement implies that information must have an
instantiation (be it ink on paper, bits in a computer's memory, or
even the neurons in a brain). Furthermore, it also implies that
information must be {\em about something}. Lines on a piece of paper,
for example, are not inherently information until it is discovered
that they correspond to something, such as (in the case of a map) to
the relative location of local streets and
buildings. Consequently, any arrangement of symbols might be viewed as
{\em potential information} (also known as {\em entropy} in
information theory), but acquires the status of information only when
its correspondence, or correlation, to other physical objects is
revealed.

In biological systems the instantiation of information is DNA, but
what is this information about?  To some extent, it is the blueprint
of an organism and thus information about its own structure. More
specifically, it is a blueprint of how to build an organism that can
best survive in its native environment, and pass on that information
to its progeny. This view corresponds essentially to Dawkins' view of
selfish genes that ``use'' their environment (including the organism
itself), for their own replication~\cite{dawkins76}.  Thus, those parts
of the genome that do correspond to something (the non-neutral
fraction, that is) correspond in fact to the environment the genome
lives in. Deutsch~\cite{deutsch97} referred to this view as ``Genes
embody knowledge about their niches''.  This environment
is extremely complex itself, and consists of the ribosomes the
messages are translated in, other chemicals and the abundance of
nutrients inside and outside the cell, the environment of the organism
proper (e.g., the oxygen abundance in the air as well as ambient
temperatures), among many others. An organism's DNA thus is not only a
``book'' about the organism, but is also a book about the environment
it lives in including the species it co-evolves with. It is
well-known that not all the symbols in an organism's DNA
correspond to something. These sections, sometimes referred to
as ``junk-DNA'', usually consist of portions of the code
that are unexpressed or untranslated (i.e., excised from the mRNA). More
modern views concede that unexpressed and untranslated regions in the
genome can have a multitude of uses, such as for example satellite DNA
near the centromere, or the poly-C polymerase intron excised from {\it
  Tetrahymena} rRNA. In the absence of a complete map of the function
of each and every base pair in the genome, how can we then decide
which stretch of code is ``about something'' (and thus contributes to
the complexity of the code) or else is entropy (i.e., random code
without function)?

A true test for whether or not a sequence is information uses the
success (fitness) of its bearer in its environment, which implies
that a sequence's information content is {\em conditional}
on the environment it is to be interpreted within~\cite{adami99}.
Accordingly, {\em Mycoplasma mycoides} for
example (which causes pneumonia-like respiratory illnesses), has a
complexity of somewhat less than one million base pairs in our nasal
passages, but close to zero complexity most everywhere else, because
it cannot survive in any other environment---meaning its genome does
not {\em correspond} to anything there.  A genetic locus that codes
for information essential to an organism's survival will be {\em
  fixed} in an adapting population because all mutations of the locus
result in the organism's inability to promulgate the tainted genome,
whereas inconsequential (neutral) sites will be randomized by the
constant mutational load.  Examining an {\em ensemble} of sequences
large enough to obtain statistically significant substitution
probabilities would thus be sufficient to separate information from
entropy in genetic codes.  The neutral sections that contribute only
to the entropy turn out to be exceedingly important for evolution to
proceed, as has been pointed out, for example, by 
Maynard Smith~\cite{maynard70b}. 

In Shannon's information theory~\cite{shannon49}, the quantity {\em
  entropy} ($H$) represents the expected number of bits required to
specify the state of a physical object given a distribution of
probabilities, that is, it measures how much information can {\it potentially}
be stored in it.

In a genome, for a site $i$ that can take on four nucleotides with
probabilities
\begin{equation}
\left\{p_C(i), p_G(i), p_A(i), p_T(i)\right\}\;, 
\end{equation}
the entropy of this site is
\begin{equation}
H_i=-\sum_j^{C,G,A,T} p_j(i)\log p_j(i)\;. \label{ent}
\end{equation}
The maximal entropy per-site (if we agree to take our logarithms to
base 4, i.e., the size of the alphabet) is 1, which occurs if all the
probabilities are all equal to 1/4. If the entropy is
measured in bits (take logarithms to base 2) the maximal entropy per
site is two bits, which naturally is also the maximal amount of
information that can be stored in a site, as entropy is just potential
information. A site stores maximal information if, in DNA, it is
perfectly conserved across an equilibrated ensemble. Then, we assign
the probability $p=1$ to one of the bases and zero to all others,
rendering $H_i=0$ for that site according to Eq.~(\ref{ent}).  The
amount of information per site is thus (see, e.g., Ref.~\cite{SSGE})
\begin{equation}
I(i)=H_{\rm max}-H_i\;. \label{siteinfo}
\end{equation}

In the following, we measure the complexity of an organism's sequence
by applying Eq.~(\ref{siteinfo}) to each site and summing over
the sites. Thus, for an organism of $\ell$ base pairs the complexity is
\begin{equation}
C=\ell-\sum_iH(i)\;.\label{complex}
\end{equation}
It should be clear that this value can only be an
approximation to the true physical complexity
of an organism's genome. In reality, sites are not independent and the
probability to find a certain base at one position may be conditional
on the probability to find another base at another position. Such {\em
  correlations} between sites are called {\em epistatic} and they can
render the entropy {\em per molecule} significantly different from the sum
of the per-site entropies~\cite{adami99}.  This entropy per molecule,
which takes into account all epistatic correlations between
sites, is defined as
\begin{equation}
H=-\sum_g p(g|E)\log p(g|E) \label{true-entropy}
\end{equation}
and involves an average over the logarithm of the conditional
probabilities $p(g|E)$ to find genotype $g$ {\em given} the current
environment $E$. In every finite population, estimating $p(g|E)$ using
the actual frequencies of the genotypes in the population (if those
could be obtained) results in corrections to Eq.~(\ref{true-entropy})
larger than the quantity itself~\cite{basharin59}, rendering the
estimate useless.  Another avenue for estimating the entropy per
molecule is the creation of mutational clones at several positions at
the same time~\cite{elena97,lenski99} to measure epistatic effects.
The latter approach is feasible within experiments with simple
ecosystems of digital organisms that we introduce in the following
section, which reveal significant epistatic effects. The technical
details of the complexity calculation including these effects are
relegated to the Appendix.

\vskip 0.25cm 
\noindent {\bf Digital Evolution} 
\vskip 0.25cm 
Experiments in evolution have traditionally been formidable due to
evolution's gradual pace in the natural world.  One successful method
uses microscopic organisms with generational times on the order of 
hours, but even this approach has difficulties; it is still impossible
to perform measurements with high precision, and the time-scale to
see significant adaptation remains weeks, at best.
Populations of {\it E.coli} introduced into new environments begin
adaptation immediately, with significant results apparent in a few
weeks~\cite{LEN1,LEN2}. Observable evolution in most organisms occurs
on time scales of at least years.

To complement such an approach,
we have developed a tool to study evolution in a computational
medium---the {\avida} platform~\cite{adami98}. The
{\avida} system hosts populations of self-replicating computer
programs in a complex and noisy environment, within a computer's
memory.  The evolution of these ``digital organisms'' is limited in
speed only by the computers used, with generations (for populations of
the order $10^3-10^4$ programs) in a typical trial taking only a few
seconds. Despite the apparent simplicity of the single-niche environment and
the limited interactions between digital organisms, very rich dynamics
can be observed in experiments with 3,600 organisms on a $60\times60$
grid with toroidal boundary conditions (see {\it Methods}).  
As this population is quite
small, we can assume that an equilibrium population will be dominated
by organisms of a single species~\cite{fn2}, whose members
all have similar
functionality and equivalent fitness (except for organisms that
lost the capability to self-replicate due to mutation).  In this
world, a new species can obtain a significant abundance only if
it has a competitive advantage (increased Malthusian parameter) thanks
to a beneficial mutation.  While the system returns to equilibrium after
the innovation, this new species will gradually exert dominance over
the population, bringing the previously-dominant species to
extinction.  This dynamics of innovation and extinction can be
monitored in detail and appears to mirror the dynamics of {\it E.
  coli} in single-niche long-term evolution
experiments~\cite{elena96}.

The complexity of an adapted digital organism according to
Eq.~(\ref{complex}) can be obtained by measuring substitution
frequencies at each instruction across the population. Such a
measurement is easiest
if genome size is constrained to be constant as is done in the
experiments reported below, though this constraint can be relaxed by
implementing a suitable alignment procedure.  In order to correctly
assess the information content of the ensemble of sequences, we need
to obtain the substitution probabilities $p_i$ at each position, which
go into the calculation of the per-site entropy Eq.~(\ref{ent}).  Care
must be taken to wait sufficiently long after an innovation, in order
to give those sites within a new species that are variable a chance to
diverge. Indeed, shortly after an innovation, previously 100\%
variable sites will appear fixed by ``hitchhiking'' on the successful
genotype, a phenomenon discussed further below.

\begin{figure}[bt]
  \centerline{\psfig{figure=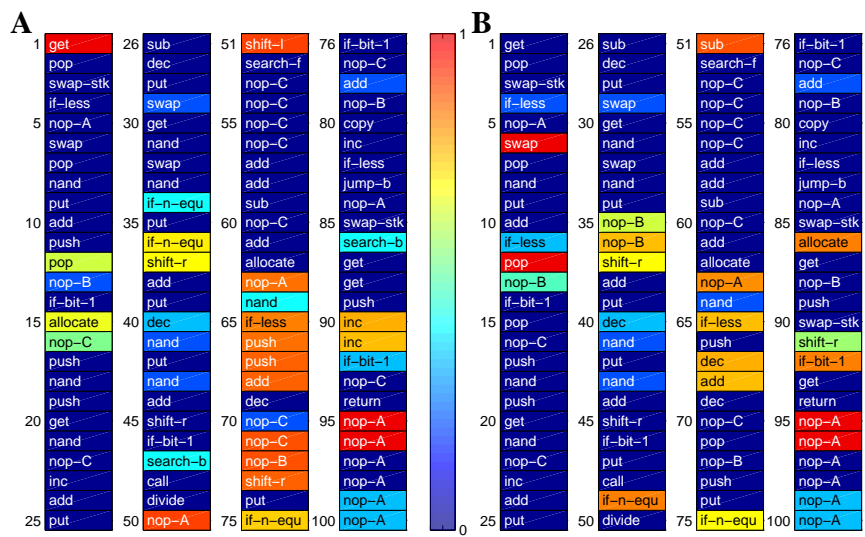,width=4.5in,angle=0}}
\caption{Typical {\avida} organisms, extracted at 2,991 (A) and 3,194 (B) 
   generations respectively into an evolutionary experiment.  Each
   site is color-coded according to the entropy of that
   site (see color bar).  Red sites are highly variable whereas blue sites are
   conserved. The organisms have been extracted just before and after
   a major evolutionary transition. }
\end{figure}

We simplify the problem of obtaining substitution probabilities for
each instruction by assuming that all mutations are either lethal,
neutral, or positive, and furthermore assume that all non-lethal
substitutions persist with equal probability. We then categorize every
possible mutation directly by creating all single-mutation genomes and
examining them independently in isolation.  In that case
Eq.~(\ref{ent}) reduces to
\begin{equation}
H_i=\log_{28}(N_\nu)\;,
\end{equation}
where $N_\nu$ is the number of non-lethal substitutions (we count
mutations that significantly reduce the fitness among the lethals). 
Note that the logarithm is taken with respect to the size of the
alphabet.

This per-site entropy is used to illustrate the variability of loci in
a genome, just before and after an evolutionary transition, in
Fig.~1.  
\vskip 0.25cm
\noindent {\bf Progression of Complexity}
Tracking the entropy of each site in the ge\-no\-me allows us to document
the growth of complexity in an evolutionary event. For example, it is
possible to measure the difference in complexity between the pair of
genomes in Fig.~1, separated by only 203 generations and a
powerful evolutionary transition.  Comparing their entropy maps, we
can immediately identify the sections of the genome that code for the
new ``gene'' that emerged in the transition---the entropy at those
sites has been drastically reduced, while the complexity increase across
the transition (taking into account epistatic effects) turns out to
be $\Delta C\approx6$, as calculated in
the Appendix. 

\begin{figure}[bt]
 \centerline{\psfig{figure=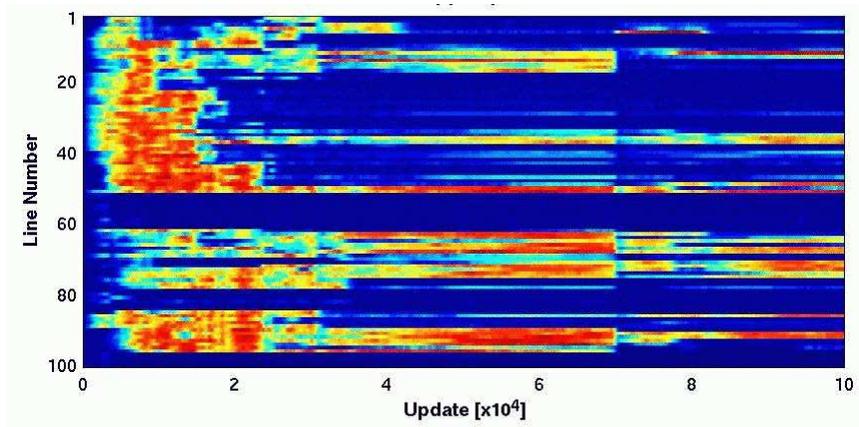,width=4.5in,angle=0}}
\caption{Progression of per-site entropy for all 100 sites throughout an
   {\avida} experiment, with time measured in ``updates'' (see {\it
     Methods}).  A generation corresponds to between 5 and 10 updates,
   depending on the gestation time of the organism.}
\end{figure}

We can extend this analysis by continually surveying the entropies of
each site during the course of an experiment.  Figure~2
does this for the experiment just discussed, but this time the
substitution probabilities are obtained by sampling the actual
population at each site.  A number of features are apparent in this
figure. First, the trend toward a ``cooling'' of the genome (i.e., to
more conserved sites) is obvious.  Second, evolutionary
transitions can be identified by vertical darkened ``bands'', which
arise because the genome instigating the transition replicates faster
than its competitors thus driving them into extinction.  As a
consequence, even random sites that are ``hitchhiking'' on the
successful gene are momentarily fixed.  

Hitchhiking is documented clearly by
plotting the sum of per-site entropies for the population (as an
approximation for the entropy of the ge\-nome)
\begin{equation}
H\approx\sum_{i=1}^\ell H(i) \label{ent-est}
\end{equation}
across the transition in Figure~3A.  By comparing this to the
fitness shown in Figure~3B, we can identify a
sharp drop in entropy followed by a slower recovery for each adaptive event
that the population undergoes.  Often, the population does not 
reach equilibrium (the state of maximum entropy given the current
conditions) before the next transition occurs.

\begin{figure}[tb!]
\centerline{\psfig{figure=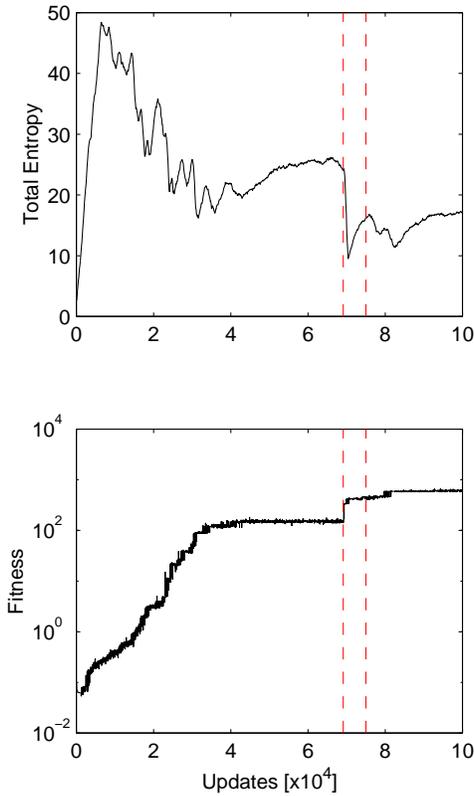,width=2.5in,angle=0}}
\caption{
(A) Total entropy per program as a function of evolutionary time.
(B) Fitness of the most abundant genotype as a function of time.
Evolutionary transitions are identified with short periods in which
the entropy drops sharply, and fitness jumps. Vertical dashed lines
indicate the moments at which the genomes in Fig.~1 A and B were
dominant.}
\end{figure}

While this entropy is not a perfect approximation of the exact entropy
per program Eq.~(\ref{true-entropy}), it reflects the disorder in the
population as a function of time.  This complexity estimate
(\ref{complex}) is shown as a function of evolutionary time for this
experiment in Figure~4.  It increases monotonically
except for the periods just after transitions, when the complexity
estimate (after overshooting the equilibrium value) settles down
according to thermodynamics' second law (see below).  This
overshooting of stable complexity is a result of the overestimate of
complexity {\em during} the transition due to the hitchhiking effect
mentioned earlier.  Its effect is also seen at the beginning of
evolution, where the population is seeded with a single genome with no
variation present.

Such a typical evolutionary history documents that the physical complexity,
measuring the amount of information coded in the sequence about its
environment, indeed steadily increases.  The circumstances under which this
is assured to happen are discussed presently.

\begin{figure}[bt]
\centerline{\psfig{figure=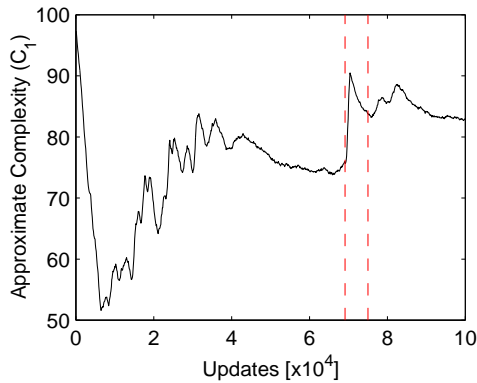,width=2.5in,angle=0}}

\caption{Complexity as a function of time, calculated according to
   Eq.~(\ref{complex}). Vertical dashed lines as in Fig.~3.} 
\end{figure}

\vskip 0.25cm
\noindent{\bf Maxwell's Demon and the Law of Increasing Complexity}
\vskip 0.25cm Let us consider an evolutionary transition like the one
connecting the genomes in Figure~1 in more detail. In this
transition, the entropy (cf. Fig.~3A) does not fully
recover after its initial drop.  The difference between the
equilibrium level before the transition and after is proportional to
the information acquired in the transition, roughly the number of
sites that were frozen.  This difference would be equal to the
acquired information if the measured entropy Eq.~(\ref{ent-est}) were
equal to the exact one given by Eq.~(\ref{true-entropy}).  For this
particular situation, in which the sequence length is fixed along with
the environment, is it possible that the complexity decreases?  The
answer is that in a sufficiently large population this cannot happen
(in smaller populations, there is a finite probability of all
organisms being mutated simultaneously, referred to as Muller's
ratchet~\cite{muller}), as a consequence of a simple application of
the second law of thermodynamics.  If we assume that a population is
at equilibrium in a fixed environment, each locus has achieved its
highest entropy given all the other sites.  Then, with genome length
fixed, the entropy can only
stay constant or decrease, implying that the complexity (being
sequence length minus entropy) can only increase.  How is a drop in
entropy commensurate with the second law?  This answer is simple also:
the second law holds only for equilibrium systems, while such a
transition is decidedly {\em not} of the equilibrium type. In fact,
each such transition is best described as a {\em measurement}, and
evolution as a series of random measurements on the environment.
Darwinian selection is a filter, allowing only informative
measurements (those increasing the ability for an organism to survive)
to be preserved.  In other words, information cannot be lost in such
an event because a mutation corrupting the information is purged due
to the corrupted genome's inferior fitness (this holds strictly for
asexual populations only).  Conversely, a mutation that corrupts the
information cannot increase the fitness, because if it did then the
population was not at equilibrium in the first place. As a
consequence, only mutations that reduce the entropy are kept while
mutations that increase it are purged.  Because the mutations can be
viewed as measurements, this is the classical behavior
of the Maxwell Demon.

What about changes in sequence length? 
In an unchanging environment, an increase or decrease in sequence length 
is always associated with an increase or decrease in the entropy, and
such changes therefore always cancel from the physical complexity, as
it is defined as the difference. 
Note, however, that while size-increasing  events do not increase
the organism's physical complexity, they are critical to continued
evolution as they provide new space (``blank tape'') to record environmental
information within the genome, and thus to allow complexity to march
ever forward.

\vskip 0.25cm
\noindent {\bf Methods}
\vskip 0.25cm For all work presented here, we use a single-niche
environment in which resources are isotropically distributed and
unlimited except for CPU time, the primary resource for this life-form.
This limitation is imposed by constraining the average {\em slice} of
CPU time executed by any genome per update to be a constant (here 30
instructions). Thus, per update a population of $N$ genomes executes
$30\times N$ instructions.  The unlimited resources are numbers that the
programs can retrieve from the environment with the right genetic
code. Computations on these numbers allow the organisms to execute
significantly larger slices of CPU time, at the expense of inferior
ones (see~\cite{adami98,lenski99}).

A normal {\avida} organism is a single genome (program) composed of a
sequence of instructions that are processed as commands to the CPU of
a virtual computer.  In standard {\avida} experiments, an organism's
genome has one of 28 possible instructions at each line. The set of
instructions (alphabet) from which an organism draws its code is
selected to avoid biasing evolution towards any particular type of
program or environment. Still, evolutionary experiments will always
show a distinct dependence on the ancestor used to initiate
experiments, and on the elements of chance and history. To minimize
these effects, trials are repeated in order to gain statistical
significance, another crucial advantage of experiments in artificial
evolution. In the present experiments, we have chosen to keep sequence
length fixed at 100 instructions, by creating a self-replicating
ancestor containing mostly non-sense code, from which all populations
are spawned. Mutations appear during the copy process, which is flawed
with a probability of error per instruction copied of 0.01. 
For more details on {\avida}, see \cite{OBA98}.

\vskip 0.25cm
\noindent {\bf Conclusions}
\vskip 0.25cm
Trends in the evolution of complexity are difficult to argue for or
against if there is no agreement on how to measure complexity. We have
proposed here to identify the complexity of genomes by the amount of
information they encode about the world in which they have evolved, a
quantity known as {\em physical complexity} that, while it can be
measured only approximately, allows quantitative statements to be made
about the evolution of genomic complexity. In particular, we show that
in fixed environments, for organisms whose fitness depends only on
their own sequence information, physical complexity must always
increase. That a genome's physical complexity must be reflected in the
{\em structural} complexity of the organism that harbors it seems to
us inevitable, as the purpose of a physically complex genome is
complex information {\em processing}, which can only be achieved by
the computer which it (the genome) creates. 

That the mechanism of the Maxwell Demon lies at the heart of the
complexity of living forms today is rendered even more plausible by
the many circumstances which may cause it to fail. First, simple
environments spawn only simple genomes. Second, changing environments
can cause a drop in physical complexity, with a commensurate loss in
(computational) function of the organism, as now meaningless genes are
shed. Third, sexual reproduction can lead to an accumulation of
deleterious mutations (strictly forbidden in asexual populations)
that can also render the Demon powerless. All such exceptions are
observed in nature. 

Notwithstanding these vagaries, we are able to observe the Demon's
operation directly in the digital world, giving rise to complex
genomes that, though poor compared to their biochemical brethren,
still stupefy us with their intricacy and an uncanny amalgam of
elegant solutions and clumsy remnants of historical contingency.  It
is in no small measure an awe before these complex programs, direct
descendants of the simplest self-replicators we ourselves wrote, that
leads us to assert that even in this view of life, spawned by and in
our digital age, there is grandeur.

\vskip 0.25cm
\noindent 
We thank A. Barr and R.E. Lenski for discussions. Access to a Beowulf
system was provided by the Center for Advanced Computation Research at
the California Institute of Technology.  This work was
supported by the National Science Foundation. 
\vskip 0.25cm
\noindent{\bf Appendix: Epistasis and Complexity}
\vskip 0.25cm
Estimating the complexity according to Eq.~(\ref{complex}) is somewhat
limited in scope, even though it may be the only practical means for
actual biological genomes for which substitution frequencies are known
(such as, for example, ensembles of tRNA sequences~\cite{adami99}).
  For digital organisms, this estimate can be sharpened by
testing {\em all} possible single and double mutants of the wild-type
for fitness, and sampling the $n$-mutants to obtain the fraction of
neutral mutants at mutational distance $n$, $w(n)$. In this manner, an
ensemble of mutants is created for a single wild-type resulting in a
much more accurate estimate of its information content.  As this
procedure involves an evaluation of fitness, it is easiest for
organisms whose survival rate is closely related to their organic
fitness, i.e., for organisms who are not ``epistatically linked'' to
other organisms in the population. Note that this is precisely the
limit in which Fisher's Theorem guarantees an increase in
complexity~\cite{maynard70a}.

For an organism of length $\ell$ with instructions taken from an
alphabet of size $D$, let $w(1)$ be the number of neutral one-point mutants 
$N_\nu(1)$ divided by the total number of possible one-point
mutations 
\begin{equation}
w(1)=\frac{N_\nu(1)}{D\ell}\;.
\end{equation} 
Note that $N_\nu(1)$ includes the wild-type $\ell$ times, for each
site is replaced (in the generation of mutants) by each of the $D$
instructions. Consequently, the worst $w(1)$ is equal to $D^{-1}$.
In the literature, $w(n)$ usually refers to the average {\em
  fitness} (normalized to the wild-type) of $n$-mutants. While this
can be obtained here in principle, for the purposes of our
information-theoretic estimate we assume that all non-neutral mutants
are non-viable~\cite{fn3}.
We have found that for digital organisms the
average $n$-mutant fitness closely mirrors the function $w(n)$
investigated here.

Other values of $w(n)$ are obtained accordingly. We define
\begin{equation}
w(2)=\frac{N_\nu(2)}{D^2\ \ell(\ell-1)/2}\;,
\end{equation}
where $N_\nu(2)$ is the number of neutral double mutants, including
the wild-type and all neutral single mutations included in $N_\nu(1)$,
and so forth.

For the genome before the transition (pictured on the left in Fig.~1)
we can collect $N_\nu(n)$ as well as $N_+(n)$ (the number of mutants that
result in {\em increased fitness}) to construct $w(n)$. In Tab. 1, we
list the fraction of neutral and positive $n-$mutants of the
wild-type, as well as the number of neutral or positive found and the
total number of mutants tried.

Note that we have sampled the mutant distribution
up to $n=8$ (where we tried $10^9$ genotypes), in order to gain statistical significance. The function
is well fit by a two-parameter ansatz
\begin{equation}
w(n)=D^{-\alpha n^\beta} \label{fit}
\end{equation}
introduced earlier~\cite{lenski99}, where
$1-\alpha$ measures the degree  of neutrality in the code
($0<\alpha<1$), and $\beta$ reflects the degree of epistasis
($\beta>1$ for synergistic deleterious mutations, $\beta<1$ for
antagonistic ones).  Using this function, the complexity of the wild-type can
be estimated as follows.

From the information-theoretic considerations in the main text, 
the information about the
environment stored in a sequence is
\begin{equation}
C=H_{\rm max}-H=\ell-H\;, \label{exact}
\end{equation}
where $H$ is the entropy of the wild-type given its environment. We
have previously approximated it by summing the per-site entropies of the
sequence, thus ignoring correlations between the sites. Using $w(n)$,
a multi-site entropy can be defined as 
\begin{equation}
H_\ell=\log_D \left[w(\ell) D^\ell\right]\;, \label{mutant}
\end{equation}
reflecting the average entropy of a sequence 
of length $\ell$. As $D^\ell$ is the total number of
different sequences of length $\ell$, $w(\ell)D^\ell$ is the number of neutral
sequences, in other words all those sequences that carry the same
information as the wild-type. The ``coarse-grained'' entropy is just
the logarithm of that number. Eq.~(\ref{mutant}) thus 
represents the entropy of a population based on one wild-type
in perfect equilibrium in an infinite population. It should
approximate the exact result Eq.~(\ref{true-entropy}) if all neutral mutants
have the same fitness and therefore the same abundance in an infinite
population. 

Naturally, $H_\ell$ is impossible to obtain for
reasonably sized genomes as the number of mutations to test in order to
obtain $w(\ell)$ is of the
order $D^\ell$.  This is precisely the reason why we chose
to approximate the entropy in Eq.~(\ref{complex}) in the first place.
However, it turns out that in most cases the constants $\alpha$ and
$\beta$ describing $w(n)$ can be estimated from the first few $n$. 
The complexity of the wild-type,
using the $\ell$-mutant entropy (\ref{mutant}) can be defined as
\begin{equation}
C_\ell=\ell- H_\ell\;.
\end{equation}
Using (\ref{fit}), we find 
\begin{equation}
C_\ell=\alpha\,\ell^\beta\;,
\end{equation}
and naturally, for the complexity based on single mutations only
(completely ignoring epistatic interactions)
\begin{equation}
C_1=\alpha \ell\;.
\end{equation}

\begin{table}[ht]
\caption{Fraction of mutations that were neutral (first column), or 
positive (second column); total number of neutral or positive genomes
found (fourth column), and total mutants examined (fifth column) as a
function of the number of mutations $n$, for the dominating genotype
before the transition.}
\vskip 0.5cm
\begin{center}
\begin{tabular}{|c|cccc|} \hline
$n$ & $N_\nu(n)$ & $N_+(n)$ & Tot. & Tried \\ \hline
1   & 0.1418 & 0.034  & 492 & 2,700 \\
2   & 0.0203 & 0.0119 & 225 & 10,000 \\
3   & 0.0028 & 0.0028 & 100 & 32,039 \\
4   & $4.6 \ 10^{-4}$ & $6.5\ 10^{-4}$ & 100 & 181,507 \\
5   & $5.7 \ 10^{-5}$ & $1.4\ 10^{-4}$ & 100 & $1.3\ 10^6$\\ 
6   & $8.6\ 10^{-6}$ & $2.9\ 10^{-5}$ & 100 & $7.3\ 10^6$\\
7   & $1.3\ 10^{-6}$ & $5.7\ 10^{-6}$ & 100 & $5.1\ 10^7$\\
8   & $1.8\ 10^{-7}$ & $1.1\ 10^{-6}$ & 34 & $1.0\ 10^9$\\
\hline
\end{tabular}
\end{center}
\end{table}
Thus, obtaining $\alpha$ and $\beta$ from a fit to $w(n)$ allows an
estimate of the complexity of digital genomes including epistatic
interactions. As an example, let us investigate the complexity
increase across the transition treated earlier. Using both neutral and
positive mutants to determine $w(n)$, a fit to the data in
Table 1 using the functional form Eq.~(\ref{fit}) yields
$\beta=0.988(8)$ ($\alpha$ is obtained
exactly via $w(1)$). This in turn leads to a complexity estimate
$C_\ell=49.4$. After the transition, we analyze the new wild-type
again and find $\beta=0.986(8)$, not significantly different from
before the transition (while we found $\beta=0.996(9)$ {\em during} the
transition).

The complexity estimate according to this fit is
$C_\ell=55.0$, leading to a complexity increase during the transition
of $\Delta C_\ell=5.7$, or about 6 instructions. Conversely, if epistatic
interactions are not taken into account, the same analysis would
suggest $\Delta C_1=6.4$, somewhat larger. The same analysis can
be carried out taking into account neutral mutations {\em only} to calculate
$w(n)$, leading to $\Delta C_\ell=3.0$ and $\Delta C_1=5.4$.

\newpage


\begin{thebibliography}{99}
\bibitem{fn2}For these asexual organisms, the species-concept is only
  loosely defined as programs that differ in genotype but only
  marginally in function.
\end{thebibliography}
\end{document}